\documentclass[12pt]{article}
\usepackage{amssymb}
\usepackage{amsmath}
\usepackage[dvips]{graphicx}
\usepackage{graphics}

\usepackage{amssymb}
\usepackage{color} 

\begin{document}

\title{Search for a Lorentz invariant velocity distribution of a relativistic gas}

\author{Evaldo M. F. Curado$^{1,2}$, Felipe T. L. Germani$^{1}$ \\
and  Ivano Dami\~ao Soares$^{1}$ \\
$^{1}$ Centro Brasileiro de Pesquisas F\'{\i}sicas -- CBPF \\
 $^{2}$ National Institute of Science and Technology \\
 for Complex Systems \\
Rio de Janeiro, Brazil 
}

\date{}

\maketitle

\abstract{
We examine the problem of the relativistic velocity distribution in a $1$-dim relativistic gas in thermal
equilibrium. We use numerical simulations of the relativistic molecular dynamics for a gas with two components,
light and heavy particles. However in order to obtain the numerical data
our treatment distinguishes two approaches in the construction of the histograms
for the {\it same} relativistic molecular dynamic simulations.
The first, largely considered in the literature, consists in constructing histograms with constant bins in the
velocity variable and the second consists in constructing histograms with constant bins in the
rapidity variable which yields Lorentz invariant  histograms, contrary to the first approach. For histograms with
constant bins in the velocity variable the numerical data are fitted accurately by the J\"uttner distribution
which is also not Lorentz invariant.
On the other hand, the numerical data obtained from histograms
constructed with constant bins in the rapidity variable, which are Lorentz
invariant, are accurately fitted by a Lorentz invariant   distribution whose derivation is
discussed in this paper. The histograms thus constructed are not fitted by the
J\"utter distribution (as they should not).
Our derivation is based on the special theory of relativity, the central limit theorem
and the Lobachevsky structure of the velocity space of the theory, where the rapidity variable
plays a crucial role. For $v^2/c^2 \ll 1$ and $1/\beta \equiv k_B T/ m_0 c^2 \ll 1$ the distribution tends to the
Maxwell-Boltzmann distribution.
}

\section{Introduction and summary of results}

In physics, it is difficult to overestimate adequately the importance
of the Maxwell-Boltzmann (MB) distribution of velocities
for gases, introduced by Maxwell in 1860 \cite{maxwell1860}. It
was the first time that a probability concept was introduced
in a physical theory, as the existing theories at the time
were purely deterministic like Newtonian mechanics and wave theory.
Actually the work of Maxwell
was the starting point for Boltzmann to elaborate his research
program on the evolution of a time dependent distribution of velocities
for gases, culminating in the articles of 1872 \cite{boltzmann1872}
and 1877\cite{boltzmann1877}, among other important papers, which are
amid the fundamental cornerstones of the modern kinetic theory of gases and of
statistical mechanics.

With the implicit use of the atomic theory of matter (at that time a
controversial theory), the new concept of entropy was established having also
as its starting point the introduction, by Maxwell, of the concept of
probability.
Since then the Maxwell-Boltzmann (MB) distribution played a fundamental role
in the statistical description of gaseous systems with a large number of constituents.
Actually, in many cases it is considerably simpler, and even as accurate as, to use the
MB distribution instead of Bose-Einstein and Fermi-Dirac distributions\cite{balian}.
However a clear limitation of the MB distribution is its nonrelativistic character,
encompassing velocities larger than the velocity of light in contradiction with
the special theory of relativity.

In the present paper we introduce a 1- dim Lorentz invariant  (LI) distribution
of velocities for a relativistic gas in thermal equilibrium which has the MB distribution
as a limit for velocities much smaller than the velocity of light
(with correspondingly relatively small temperatures). A discussion about
this distribution in the 1-dim and 3-dim cases was given by two of us in \cite{evaldo-ivano}.
Our derivation was based on three pillars: the
special theory of relativity, the central limit theorem and the fact that the velocity space of special
relativity is a Lobachevsky space, where the additivity of
velocities in the Galilean relativity is transferred to the additivity of rapidities.
\begin{figure*}
\begin{center}\includegraphics*[height=4cm,width=6cm]{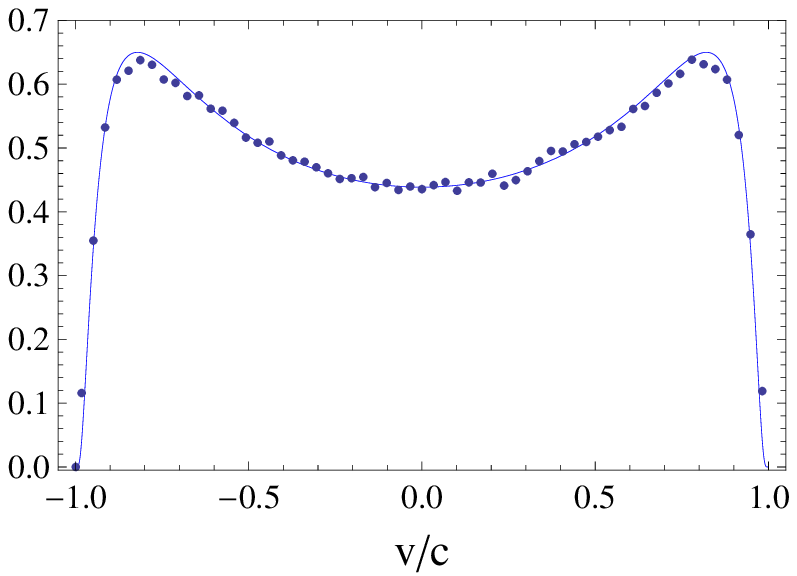}
\includegraphics*[height=4cm,width=6cm]{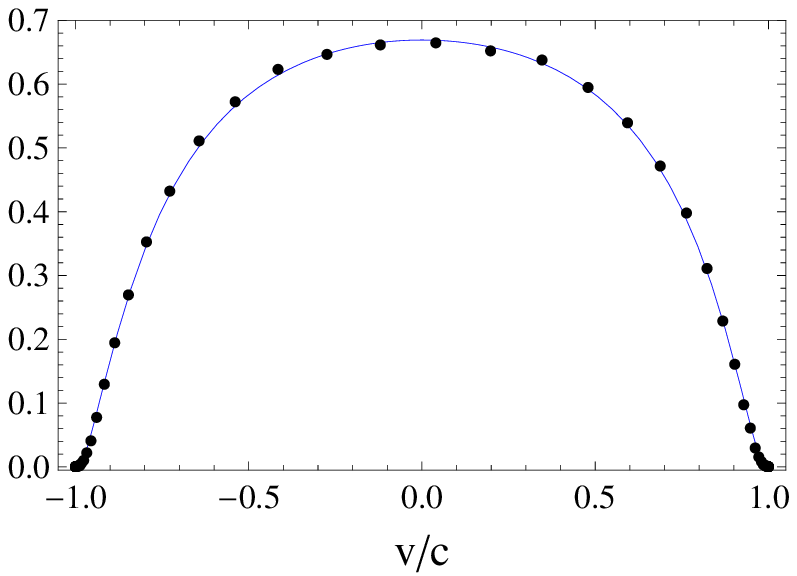}
\caption{Plot of the equilibrium velocity distributions of heavy particles ($N_2=2,500$) obtained
from the {\it same numerical simulation} for the relativistic molecular dynamics of a 1-dim gas with $N=5,000$ particles:
({\it left}) the points are obtained from a histogram
of constant velocity bins, which is not LI, and the continuous curve is the best fit
of the points to the J\"uttner distribution (\ref{juttner}), with best
fit parameter $m_0 \beta_J \simeq 0.8570$ (rms error $\simeq 0.017$).
({\it right}) The points are obtained from a histogram of constant rapidity bins, which is LI,
transformed into a normalized histogram in the velocities (via eqs. (\ref{w}) and (\ref{rap})) with non
constant bins but still LI. The continuous curve is the best fit of the points to the LI velocity distribution (\ref{eq4}) with best fit parameter $\beta \simeq 1.4070$ (rms error $\simeq 0.004$).}
\label{figdist22}
\end{center}
\end{figure*}

 We have made a numerical simulation of the relativistic dynamics for a 1-dim relativistic molecular gas
with two components (light and heavy particles) in the same
vein of Cubero et al. \cite{hanggi2007}. We used $N_1=2,500$ light particles of mass $m_0$ and $N_2=2,500$ heavy
particles of mass $2 m_0$, with a total of $100$ relativistic simulations (taking the box as the frame
of reference in the simulations),
and collected the data fixing the number $60$ of points to be obtained
in the histograms. The initial random distributions of the velocity of the particles must satisfy $-1< v_i/c <1$
with $\sum_i v_i=0$ (so that the box has zero velocity). The control of how much relativistic is the
system is made by how the interval of the initial random
distributions is closer to $1$: the random choice of $|v_i/c|$ in the interval $[0.75,1)$ is considered
less relativistic than in the interval $[0.85,1)$. The four-momentum $P_a$ of the particles is
conserved for the whole relativistic evolution, with the relative
speed of the colliding particles being a constant in the collision. The numerical methods follow closely
\cite{allen,haile,dunkel}.

To proceed, for the {\it same} relativistic simulations we adopted
two distinct approaches in the construction of the histograms of the
distribution of velocities in the thermalization limit:
the first, histograms with constant velocity bins (which are not LI)
and the second, histograms with constant rapidity bins (which are LI).
In the second approach, from the constant rapidity histograms we obtain  histograms of the velocities
by using the inverse transformations
of Eqs. (\ref{rap}) and (\ref{w}).
These transformations change the relative size of
the bins, with the bins no longer equally spaced in a velocity
scale as they were in a rapidity scale. However these histograms still remain LI.

It is crucial to remark here that the use of the rapidity is not merely a change of
variables. The core of our new approach in this paper is that we treat the numerical data
obtained from the molecular dynamics simulations by constructing the histograms with constant bins
in the rapidity which is a LI procedure. This yields a LI probability distribution that
has a completely distinct measure from previous treatments in the literature (cf. \cite{hanggi2007})
where the construction of histograms is made with constant bins in the velocity, a procedure
that is clearly not LI and results in a completely distinct distribution. The distribution 
made with constant bins in the velocity cannot be
transformed into the LI distribution obtained in this paper by a change of variables.

\begin{figure*}
\begin{center}\includegraphics*[height=3.8cm,width=6cm]{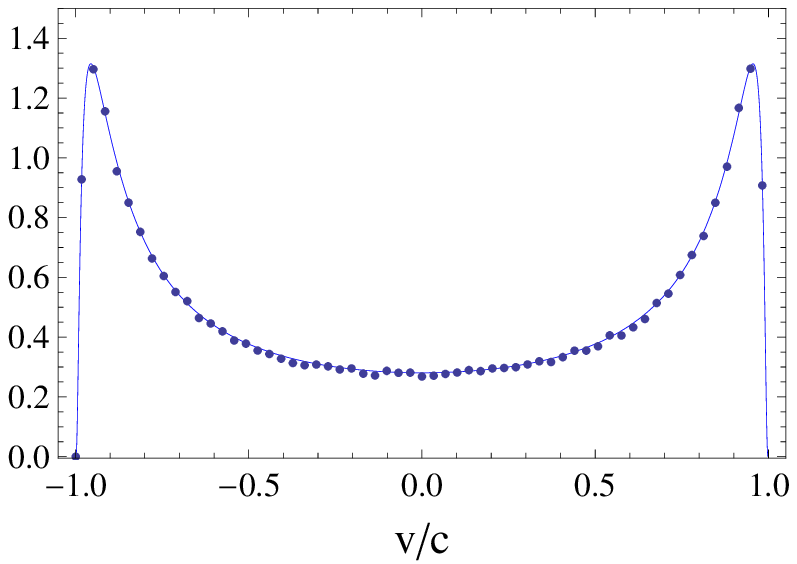}
\includegraphics*[height=4cm,width=6cm]{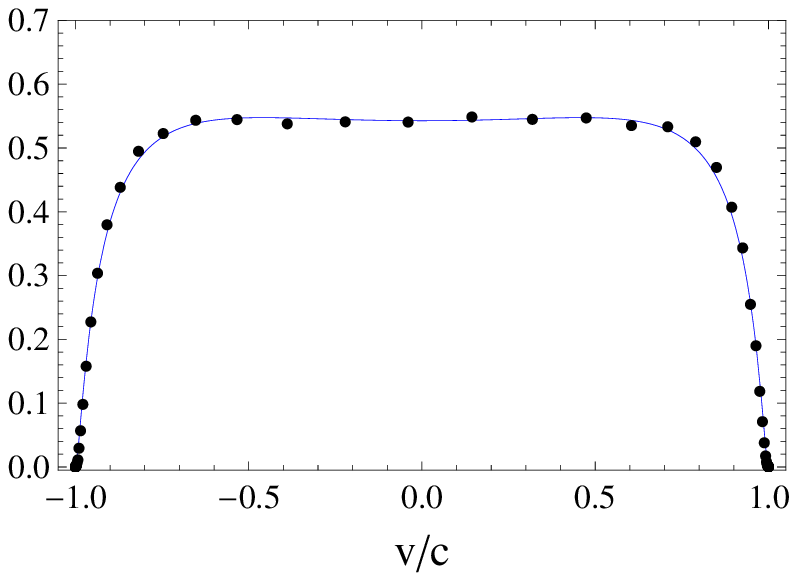}
\caption{Plot of the equilibrium velocity distributions of light particles ($N_1=2,500$) corresponding
to the {\it same numerical simulation} of Figs. \ref{figdist22}:
({\it left}) the points are obtained from a histogram
with constant velocity bins, which is not LI, and the continuous curve is the best fit
of the points to the J\"uttner distribution (\ref{juttner}), with best fit
parameter $m_0\beta_J \simeq 0.8862$ (rms error $\simeq 0.009$).
({\it right}) The points are obtained from a histogram of constant rapidity bins, which is LI,
transformed into a normalized histogram in the velocities (via eqs. (\ref{w}) and (\ref{rap})) with non
constant bins but still LI. The continuous curve is the best fit of the points to the Lorentz
invariant velocity distribution (\ref{eq4}) with best fit parameter $\beta \simeq 0.9254$ (rms error $\simeq 0.008$).}
\label{figdist23}
\end{center}
\end{figure*}

Part of our main results are displayed in Figs. \ref{figdist22} and \ref{figdist23}
where we show the equilibrium velocity distributions
derived from our numerical simulations for the light and heavy components of the 1-dim relativistic gas,
respectively. For the same simulation, the points are obtained either from a histogram of
constant velocity bins (left figures), which is not LI, or from
a histogram of constant rapidity bins (right figures) which is LI.

 In Figs. \ref{figdist22} and \ref{figdist23} ({\it left}), where histograms of constant velocity bins are adopted,
the continuous curves correspond to the best fit of the J\"uttner distribution (\ref{juttner}) \cite{juttner1911}.
The accurate fits reproduce a result of Cubero et al. \cite{hanggi2007} which is a
standard reference in the literature on the subject, being considered as a clear evidence
that J\"uttner's is the correct relativistic distribution although
not being a LI distribution. With the same procedure this result was later also
verified for the case of 2-dim\cite{iranianos,iranianos1} and 3-dim\cite{dunkel1} distributions.

 In Figs. \ref{figdist22} and \ref{figdist23} ({\it right}), where histograms of constant rapidity bins
are adopted (properly transformed to a histogram where the abscissa is the difference of velocities),
the continuous curves correspond to the best fit of the LI
velocity distribution (\ref{eq4}).

 These two contrasting results, derived from the same numerical simulations, are connected
to the distinct particular choices made on constructing histograms, namely, whether we choose the histogram to
be LI or not, a fact that has not yet been taken into account in the literature.

\begin{figure}
\begin{center}
\includegraphics*[height=4cm,width=6cm]{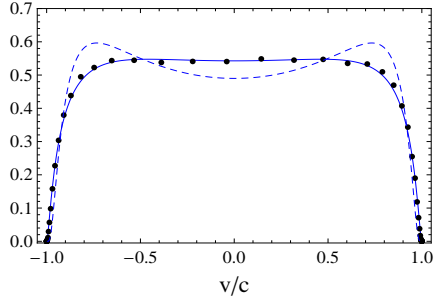}
\caption{Plot of the velocity distribution of Figs. \ref{figdist23} ({\it right}):
the {\it dashed curve} is the best fit of the points to the J\"uttner distribution (\ref{juttner}), with
best fit parameter $m_0\beta_J  \simeq 2.0406$ (rms error $\simeq 0.04$).
We see that while the LI velocity distribution fits quite accurately the points,
the J\"uttner distribution cannot be a candidate for a LI distribution.}
\label{fig2}
\end{center}
\end{figure}

 On the other hand the graphs of Fig. \ref{fig2}, done for light particles, show that the points obtained from a histogram
of constant rapidity bins (which is LI) fit quite well the LI distribution
(\ref{eq4}) (continuous line) and give evidence that the J\"uttner distribution (\ref{juttner}) cannot
be a candidate for a LI velocity distribution (dashed line).

\section{A proposal for a 1-dim LI velocity distribution}

We start by presenting a derivation of the MB velocity distribution which differs from the
derivation used by Maxwell but which will be
more appropriate
in our derivation
of a LI velocity distribution. Let us first consider the addition of velocities in the
Galilean space. We know that the velocities add according the rule
${\bf v} =  \sum_i {\bf v}_i $. Assuming that the velocities of the particles ${\bf v}_i$ are
random variables, with zero mean, and considering that the sum is over
a very large number of particles, we have - by the central limit theorem \cite{central} - that
the probability distribution of velocities for the random variable
${\bf\nu} \equiv (1/\sqrt{N}) \sum_i {\bf v}_i$ approaches the distribution
$P({\bf \nu}) \propto \exp(-b {\bf \nu}^2)$ or
$P({\bf v}) \propto \exp(-b {\bf v}^2/N)$,
recovering thus the famous MB distribution if we plot $\sqrt{N} P({\bf v})$ versus ${\bf v/\sqrt{N}}$.

As well known, in the special theory of relativity the
relative velocity ${\bf v}$ of two particles with
arbitrary velocities ${\bf v_1}$ and ${\bf v_2}$, with respect to a fixed inertial frame, is given by
{\small
\begin{eqnarray}
\label{eq5}
{\bf v}=\frac{{\bf v_1}-{\bf v_2}+(\gamma(v_2)-1)({\bf v_2}/v_2^2)[{\bf v_1}\cdot{\bf v_2}-v_2^2]}{\gamma(v_2)(1-{\bf v_1}\cdot {\bf v_2}/c^2)}.
\end{eqnarray}
}
The above expression also holds with the interchange $\mathbf{v_1} \leftrightarrow \mathbf{v_2}$.
The square of the modulus of the relative velocity is given by
{\small
\begin{eqnarray}
\label{eq6}
v^2=\frac{({\bf v_1}-{\bf v_2})^2 - (1/c^2)[{\bf v_1}\wedge{\bf v_2}]^2}{(1-{\bf v_1}\cdot {\bf v_2}/c^2)^2},
\end{eqnarray}
}
which is symmetric with respect to {\small${\bf v_1}$} and {\small ${\bf v_2}$}. It is also a well known fact that
the square of the relative velocity (\ref{eq6}) is invariant under Lorentz transformations\cite{fock}.
Let us now consider a fixed inertial reference frame,
say the laboratory frame.
For simplicity, we will assume one dimensional
only. Let the velocity of two particles be $v_1$ and $v_2$, as measured in this
inertial frame. The relative velocity of the two particles is given from (\ref{eq6}) as
{\small
\begin{eqnarray}
\label{eq1}
v=\frac{v_1 + v_2}{1 + v_1 v_2/c^2} \,
\end{eqnarray}
}
which is invariant under Lorentz transformations.
However (\ref{eq1}) can be rewritten as
{\small
\begin{eqnarray}
\label{eq2}
\frac{1 + v/c}{1 - v/c}=\Big( \frac{1 + v_1/c}{1 - v_1/c}\Big) \Big(\frac{1 + v_2/c}{1 - v_2/c}\Big),
\end{eqnarray}
}
and can be extended to any number of particles,
{\small
\begin{eqnarray}
\label{eq3}
\frac{1 + v/c}{1- v/c}=\prod_{i}\Big( \frac{1 + v_i/c}{1 - v_i/c}\Big).
\end{eqnarray}
}
Taking the logarithm on both sides of Eq. (\ref{eq3}) and defining
{\small
\begin{eqnarray}
\label{si}
\sigma_i \equiv \frac{1}{2} \ln \left(\frac{1 + v_i/c}{1 - v_i/c} \right)
= \tanh^{-1}(v_i/c) \,\,,~~\sigma_i \in (-\infty,\infty),
\end{eqnarray}
}
we can express (\ref{eq3}) as
{\small
\begin{eqnarray}
\label{ssoma}
\sigma= \sum_i \sigma_i            \, .
\end{eqnarray}
}
The variable defined in (\ref{si}) is the rapidity associated with the velocity $v_i$ and
will play a fundamental role in the remaining of the paper. Accordingly the rapidity of
relatives velocities are therefore quantities which are LI and are additive
in the arithmetic sense.

Let us then consider a relativistic gas with a large number of particles, each with
a velocity $v_i$ as measured with respect to the inertial frame of the laboratory\footnote{In fact any inertial frame can be considered with respect to which the
velocities are constructed. The results of the numerical simulations and the related histograms
will not change as can be verified.}.
We concur that any Boltzmann-type equation (relativistic or not)
that give rise to a universal stationary velocity distribution implicitly assumes the presence of a spatial
confinement, thus singling out a preferred frame (cf. Cubero et al.\cite{hanggi2007} and references therein).
However as we will see the LI distribution function derived here will be independent of
such a singling out of frame, as the variables involved are LI.
If we assume that the velocities $v_i$ are independent and random variables, with zero mean,
the variables $\sigma_i$'s are also independent and random
with zero mean, and we have -- in accordance with the central limit theorem -- that the probability distribution
for the variable $s \equiv \sigma/\sqrt{N}$ in an interval $s$ and $s+ds$ approaches
$P(s)ds =  C_1 e^{-\tilde{\beta} s^2} ds$
or, denoting $\beta \equiv \tilde{\beta}/N$,
$ P(\sigma)d\sigma =  C_1 e^{-\beta \sigma^2} d\sigma / \sqrt{N}$,
where $C_1=\sqrt{\tilde{\beta}/\pi}$ is a normalization constant.
\noindent Using (\ref{si}) and that $d \sigma = \gamma^2 dv/c$, where
{\small $\gamma = 1/\sqrt{1-v^2/c^2}$}, we can write the probability
distribution for the velocities of a 1-dim relativistic gas as
{\small
\begin{eqnarray}
\label{eq4}
P(v)dv= {C_1} e^{-\beta \Big( \tanh^{-1}(v/c) \Big)^2} \gamma^2 dv / c \sqrt{N}.
\end{eqnarray}
}
The factor {\small $\gamma^2 dv$} in (\ref{eq4}) is the LI 1-dim line element.
The adimensional parameter $\beta=m_0 c^2/\kappa_B T$, where $\kappa_B$ is the Boltzmann constant,
decreases with the temperature $T$. Such temperature is obviously LI.
The distribution (\ref{eq4}) is a 1-dim LI velocity
distribution for a relativistic gas.
\begin{figure}
\begin{center}
\includegraphics*[height=4cm,width=6cm]{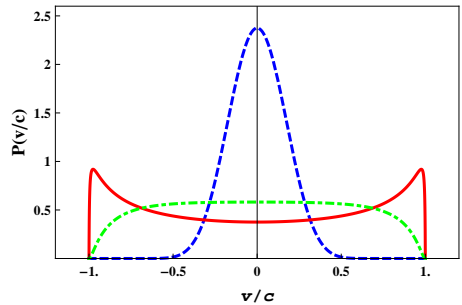}
\caption{Plot of the 1D LI velocity distribution (\ref{eq4})
for a fixed mass and decreasing values of the adimensional parameter $\beta$ (successively dashed, dashdotted
and continuous curves, cf. text). The distribution is zero for $v^2/c^2 \geq 1$,
as expected.}
\label{figdist1}
\end{center}
\end{figure}
In Fig. \ref{figdist1} we plot the distribution (\ref{eq4}) for a fixed mass and several decreasing
parameters $\beta$.

 The LI velocity distribution changes the concavity at zero velocity for
the particular value $\beta = 1$ which separates highly relativistic domains from  domains that
contain the Maxwell-Boltzmann limit.
Above $\beta=1$ the zero velocity concavity is negative; below $\beta=1$ the concavity is positive characterizing
an extreme relativistic limit for $\beta \ll 1$. On the opposite side for $\beta \gg 1$ we have approach the MB regime.
For the J\"utnner distribution (\ref{juttner}) with $d=1$
the concavity at zero velocity changes for $m_0 \beta_J=3$ showing that
a high relativistic regime is attained at a lower temperature, as compared to the LI distribution (\ref{eq4}).

 We should finally remark that the variable $\sigma$ is in fact a particular case of the LI
distance measure in the 3-dim Lobachevsky velocity space, which is the space of velocities
in the special theory of relativity, as discussed in \cite{evaldo-ivano}.

\section{Discussion on how to construct histograms of velocities for a relativistic gas\label{sectionIII}}

In recent past years a large number of numerical simulations of a relativistic gas has appeared
in the literature \cite{hanggi2007, hanggi2008,iranianos,iranianos1,dunkel1,hans2011}. In these
simulations the authors have studied some proposals of distribution of velocities for a relativistic gas,  like
the J\"uttner distribution \cite{juttner1911} or the J\"uttner modified distribution \cite{hanggi2007}.
The overall conclusion of all of these papers is that the
J\"uttner distribution matches accurately the data obtained by relativistic molecular dynamic simulations,
convincing the majority of the scientists that J\"uttner is the correct relativistic distribution.
 In the two following subsections we discuss some issues connected to the Lorentz invariance in these simulations
and we distinguish approaches to construct the histograms of velocities to be compared with the theoretical distributions.

\subsection{Histograms with constant velocity bins }

In the molecular dynamical simulations used previously in the above cited papers,
the histograms of velocities have been constructed using the {\it event driven simulation} method \cite{allen,haile}.
After starting the simulations with the particles of the gas having random position and velocities
and waiting the equilibration time, they collect the velocities of the particles at a precise moment
and store them in a pool.
The procedure is repeated many times and when the pool of velocities has
a large number of data a histogram is constructed
having in the abscissa equal size bins of difference of velocities.
In our simulations here we have $100$ samples, each with $N=5,000$ particles
and equilibration time equal to $100 \times N$.

Let us discuss the construction of these histograms in one dimension.
The extension for two or three dimensions is straightforward.
In one dimension we divide -- in the abscissa -- the domain of the values of the velocities obtained, say $[-c, c]$,
in $n$ intervals (bins) of equal size ($n=60$). In the ordinate of the $j$-th bin ($j=1,\cdots, n$)
we put the number of all particles having, at the moment of the measurements,
velocities comprised between the limits of the bin.
The histogram thus obtained, after normalized, is compared with the theoretical relativistic distributions proposed in the literature, as the
J\"uttner distribution, that in $d$-dimension is given as
\begin{equation}
\label{juttner}
f_J(v,m,\beta_J)=\frac{1}{Z_J} m_0^d \, \gamma({\bf v})^{2+d} \, \exp[-\beta_J m_0 \gamma({\bf v})] \, ,
\end{equation}
where $m_0$ is the rest mass of the particles and $\beta_J=c^2/\kappa_B T$, where $\kappa_B$ is the Boltzmann constant.
The one dimensional J\"uttner distribution is obtained by simply taking $d=1$ in
Eq. (\ref{juttner})\footnote{In the 1-dim case it is necessary that a fraction of the $N$ particles considered in the simulations
(let us say $N/2$) have distinct masses in order that the
distribution of initial velocities do not remain unaltered and that the system
undergo equilibration\cite{dunkel,hanggi2007}.}.
Adopting this procedure, all the numerical simulations in the literature referred to here match accurately
with the J\"uttner distribution (cf. for instance \cite{hanggi2007} in 1-dim, \cite{iranianos,iranianos1} in 2-dim
and \cite{dunkel1} in 3-dim).
As a consequence of these results the present well-established position adopted by most of the scientists
of the statistical mechanics community is that the correct
relativistic distribution is the J\"uttner distribution. A striking illustration of this can be seen in
\cite{hanggi2007} (Figure 1), \cite{iranianos} (Figure 1) and \cite{dunkel1} (Figure 1).

 In our simulations  we have also performed a 1-dim relativistic dynamical molecular simulation
and the comparison of the histogram (constructed with constant velocity bins) with the J\"uttner
distribution is displayed in Figures \ref{figdist22} and \ref{figdist23} ({\it left}) for heavy and light  particles.
As it can be seen, the
agreement is also quite accurate.

 However we consider that two points are not satisfactory  within this approach. The first one is that the
theoretical J\"uttner distribution {\it is not} a LI distribution. In fact not only the relativistic energy,
present in the argument of the exponential of Eq. (\ref{juttner}), is not a LI quantity but
also the correct $\gamma({\bf v})$ factor needed to guarantee Lorentz invariance in 3D is $\gamma^4({\bf v})$
(connected to the invariant element of volume of the Lobachevsky space of relativstic velocities) and not $\gamma^5({\bf v})$
as it appears in the J\"uttner distribution.
The second point is that, in the construction of the histograms, in the abscissa, where the $\mu$-th bin
(which initially is constant for any $\mu$ and equal to $\Delta = 2c /n $, for $n$ bins)
corresponds to particles having velocities going from $v_{\mu-1}$ to $v_\mu$,
is not LI,  since the difference
$\Delta_\mu=v_{\mu}- v_{\mu-1}$ (the size of the $\mu$-th bin) changes
 under a Lorentz transformation, i.e., all the
bins in the abscissa change differently
under a Lorentz transformation. Therefore, the diagram is not
LI as well, presenting a different form in each reference frame.
Consequently, we have the following (uncomfortable) situation: a histogram that is not LI matches
well a distribution  of velocities (J\"uttner) that is not as well LI. This certainly is not a
satisfactory theoretical framework. What we have to search is to construct a LI histogram of
velocities and then try to find a LI distribution of velocities that matches it. This is the aim of our work.

 We would like to remark that in a non-relativistic gas, whose particles obey the Maxwell-Boltzmann distribution,
the whole system should be invariant with respect to a Galilean transformation of velocities.
In a one dimension gas, for simplicity, let us consider the difference $\nu_i$ of velocities of the $i$-th particle ($i = 1, \cdots, N$)
with the velocity of the box in a Galilean inertial frame, $\nu_i \equiv v_{i} -  v_{\rm box}$. In the abscissa let us divide the
domain of the values of the variable $\nu_i$ obtained in the simulations
(say $[-{\nu}_{\rm min}, {\nu}_{\rm max}]$ $\subset$ $(-\infty,\infty)$) in $n$ intervals (bins) of equal size where the
$\mu$-th bin ($\mu=1,2, \cdots n$) is written $\Delta_{\mu}={\nu}_{\mu}-{\nu}_{\mu-1}$ ($ \equiv \Delta$ for all $\mu$), with $\nu_0 \equiv -\nu_{min}$ and
$\nu_\mu = \mu \Delta + \nu_0$.
The size of each bin is equal to $({\nu}_{\rm min}+{\nu}_{\rm max})/n$,
 where $\nu_{min}$ is the maximum  velocity obtained in direction $-x$ and $\nu_{max}$ is the maximum velocity obtained in direction $+x$.
We remark that $\Delta_\mu$,  for any $\mu$, remains constant after a Galilean transformation.
If in the ordinate we put in the $\mu$-th bin the number of particles whose associated  value of
$\nu_{i} = v_{i} -  v_{\rm box}$ is comprised between ${\nu}_{\mu-1}$ and ${\nu}_{\mu}$,
the ordinate also does not change after a Galilean transformation. This implies that the whole
histogram thus constructed
is Galilean invariant and is well-fitted
with a Gaussian centered at zero in any Galilean frame. This is the framework we want to
reproduce in the relativistic case.

 We also note that if, in the ordinate, we put in the $\mu$-th bin the number of particles
having velocities comprised between $ v_{\mu}$ and $v_{\mu-1}$ (not the difference of velocity
of the particle and the velocity of the box as previously),
after a Galilean transformation the
histogram fits well with a Gaussian centered at a nonzero velocity, that is the velocity associated with the Galilean transformation. Therefore, the best way to construct a histogram that is Galilean invariant
is considering the difference of velocities between the particles and the box. The histogram so constructed fits with a Gaussian centered at zero in any Galilean frame.

\subsection{LI histograms: constant rapidity bins\label{sectionIIIB}}

In the relativistic molecular dynamics, we also use the {\it event driven simulation} method
with the relativistic scattering rules among the particles. In order to construct a histogram
that is relativistic, or Lorentz, invariant, following the scheme of the non-relativistic case
(that is Galilean invariant), we want that nor the abscissa neither the ordinate of the histogram change after a Lorentz transformation.
Clearly, the bins constituted by the difference of velocities $\Delta_\mu=\nu_{\mu} - \nu_{\mu-1}$ are not LI.
It is therefore crucial to use in the abscissa a quantity that is LI.
To do this, let us first consider the relativistic difference $\mathbf{\mathfrak{v}}_i$ between the
velocity of the $i$-particle ($\mathbf{ v}_i$) and the velocity of the box ($ \mathbf{\mathfrak{ v}}_{box}$), given by (cf. (\ref{eq6}))
\begin{equation}
\label{w}
{\mathfrak{v}_i}^2 = \frac{({\bf v}_i - {\bf v}_{box})^2}{(1-{\bf v}_i \cdot {\bf v}_{box}/c^2)^2} \, .
\end{equation}
We now consider the rapidity, $s_i$, of this difference, concerning the particle $i$,
that is a LI quantity,
\begin{equation}
\label{rap}
s_i = \frac{1}{2}\ln(\frac{1+\mathfrak{v}_i/c}{1-\mathfrak{v}_i/c})= \tanh^{-1} (\mathfrak{v}_i/c) \, .
\end{equation}
We then propose to construct the abscissa of the histogram using
the values of the rapidity, that now goes from minus to plus infinity.
In the simulations we take the minimum and maximum numerical values obtained  for
the rapidity in all samples and divide  this interval in the abscissa in
 equal rapidity-bins,
where in the $\mu$-th rapidity-bin we put all particles having values of $\{s_i\}$ going from
the inferior and superior limits of the bin.
The normalized histogram constructed in this way is {\it invariant by a Lorentz transformation}, as expected,
but it is a histogram of the rapidity. In order to get a histogram of the velocities, or of the difference of
velocities between the particle and the box, we use the inverse transformations of Eqs. (\ref{rap}) and \eqref{w}
(once we know the velocity of the box) on the abscissa.  These transformations will change the relative size of the bins,
since now the bins are no longer equally spaced in a velocity
scale, as they are in a rapidity scale. However, even if we now represent the
rapidity histogram in a velocity-scale, this histogram is still LI.

In Figure \ref{fig2}  we show the histogram in a scale of difference of velocities
for $N=2,500$ light particles
 (the simulations have also $N=2,500$ heavy particles)
obtained from the histogram of the rapidities for the light particles.
The agreement of the numerical data with the best fit of the theoretical LI velocity
distribution (\ref{eq4}) is excellent (continuous line), with a normalized rms between the curve and the points $\simeq 0.008$. We also represent in Figure \ref{fig2}  the best-fit of the
J\"uttner distribution (dashed line) with the data of the histogram, where the parameter $m_0 \beta_{J}$ has been adjusted. Clearly
the J\"uttner distribution does not fit the data, in contrast to the case where the histograms are
constructed, from the beginning, with constant bins of velocities. This indicates that the J\"uttner distribution cannot be
considered as the relativistic partner of the Maxwell-Boltzmann distribution.
We would like to remark that in our simulations we obtained, with the method of
constant bins of velocities, a good agreement with the results of \cite{hanggi2007},
matching the Juttner distribution.
Our results were obtained from $100$ simulations with $N=5,000$ particles,
with a simulation time of $T=100 \times N$
and with a fixed number $60$ of event points for the equilibrium configuration.
The best fit of the J\"uttner distributions yields values of the parameter $m_0\beta_J$ with a relative difference
of only $3\%$ for heavy and light particles (cf. Figs. \ref{figdist22}-\ref{figdist23} ({\it left})).
However, larger differences of the values of $\beta$ appear in the LI distributions  and we are
performing further simulations to completely clarify this point.

\section{Conclusions}

We have examined here the full relativistic molecular dynamics of a 1-dim gas, by basically reproducing the
numerical simulations made in a landmark paper on the subject\cite{hanggi2007}, where the authors
established numerically that the J\"uttner velocity distribution function is the correct
generalization of Maxwell's velocity distribution in special relativity. We were led to look
into this problem again since we considered such conclusion not satisfactory mainly due to the fact
that the J\"uttner distribution is not LI. We therefore approached this problem using
the same numerical simulations but adopting two distinct procedures on constructing the histograms
with the data obtained. The first one corresponds to the analysis of \cite{hanggi2007}, where
the histograms are constructed with equal size bins in the velocity variables and is -- by
construction -- not LI. Using this procedure we were able to reproduce the results
of \cite{hanggi2007}, where the equilibrium distribution of the velocities resulting from the relativistic
simulations fits accurately the J\"uttner distribution.
The second procedure, which is a new contribution
of this paper, adopts histograms based on the rapidity, a variable which is
LI and is associated with the relative velocity of the particles: the histograms are
constructed with {\it equal size bins in the rapidity} and are LI in the same way that
the histograms of Maxwell-Boltzmann distributions, constructed with {\it equal size bins in the velocity},
are Galilean invariant. From these rapidity histograms we get a histogram of the relative velocities, or of the difference of
velocities between the particle and the box, by using the inverse transformations of Eqs. (\ref{rap}) and \eqref{w}
(once we know the velocity of the box) on the abscissa.  These transformations will change the relative size of the bins,
since now the bins are no longer equally spaced in a velocity
scale, as they are in a rapidity scale. However, even if we now represent the
rapidity histogram in a velocity-scale, this histogram is still LI.
We then show that the equilibrium distribution of velocities is accurately fitted by the
LI velocity distribution  (\ref{eq4}) proposed in this paper but not by the J\"uttner distribution.
In this way we believe that the symmetric connection between the Maxwell-Boltzmann velocity distribution
and the LI velocity distribution (\ref{eq4}) is established,
the first Galilean invariant and the second LI, by using the variable {\it relative velocity
with respect to a given inertial frame}, provided that the evaluation of relative velocities
involves respectively the ``arithmetic addition'' or the ``Lobatchevsky addition''.

\section*{acknowledgments}

We are grateful to Prof. Constantino Tsallis for stimulating
discussions and suggestions that were fundamental to the development of this paper. We also
acknowledge the Brazilian scientific agencies CNPq, FAPERJ and CAPES for financial support.


\begin{thebibliography}{0}

\bibitem{maxwell1860} J. C. Maxwell, Philosophical Magazine XIX (1860) 19-32 and XX (1860) 21-37.

\bibitem{boltzmann1872} L. Boltzmann, Wiener Berichte 66 (1872) 275-370.

\bibitem{boltzmann1877} L. Boltzmann, Wiener Berichte 76 (1877) 373-435.

\bibitem{balian} R. Balian, {\it From Microphysics to Macrophysics}: Methods and Applications of Statistical Mechanics, Vol. I, Springer Verlag (Berlin, 1991).

\bibitem{evaldo-ivano} Evaldo Curado and Ivano Dami\~ao Soares, A LI velocity distribution
for a relativistic gas, arXiv:1406.0777 (2014).

\bibitem{hanggi2007}  D. Cubero, J. Casado-Pascual, J. Dunkel, P. Talkner and P. Hanggi, Phys. Rev. Lett. 99 (2007) 170601.

\bibitem{allen} M. P. Allen and D. J. Tildesley, {\it Computer Simulations of Liquids} (Clarendon Press, Oxford, 1991).

\bibitem{haile} J. M. Haile, {\it Molecular Dynamics Simulations: Elementary Methods} (Wiley, New York, 1992).

\bibitem{dunkel} J. Dunkel and P. H\"anggi, Physica A 374, 559 (2007).

\bibitem{juttner1911} F. J\"uttner, Ann. Phys. (Leipzig) 18 (1911) 856.

\bibitem{iranianos} A. Montakhab, J. Casado-Pascual and M. Barati, Phys. Rev. E 79, 031124 (2009).

\bibitem{iranianos1} M. Ghodrat, A. Montakhab, Comp. Phys. Comm. 182 (2011) 1909.

\bibitem{dunkel1} J. Dunkel,P. H\"anggi and S. Hilbert, Nat. Phys, 5, 741 (2009).

\bibitem{central} L. E. Reichl, {\it A Modern Course in Statistical Mechanics}, John Wiley \& Sons (New York, 1998).

\bibitem{fock} V. Fock, {\it The Theory of Space, Time and Gravitation}, Pergamon Press (Oxford, 1964), Chapter I, Sec. 17.

\bibitem{hanggi2008}  F. Debbasch, Physica A 387 (2008) 2443.

\bibitem{hans2011} M. Mendoza, N. A. Araujo, S. Succi and H. J. Herrmann, Scientific Rep. 2 (2011) 611.


\end{thebibliography}
\end{document}